\documentclass[onecolumn,amsmath,amssymb]{qm2008_abs}
\usepackage{graphicx}
\usepackage{dcolumn}
\usepackage{bm}
\usepackage{subfigure}

\def \etal{{\sl et al.~\/}}
\def \ibid{{\sl ibid~\/}}
\def \np{{\sl Nucl.\ Phys.~\/}}
\def \pl{{\sl Phys.\ Lett.~\/}}
\def \pr{{\sl Phys.\ Rev.~\/}}

\begin{document}

\title{{\Large QGP Susceptibilities from PNJL Model }}

\bigskip
\bigskip
\author{\large Sanjay K. Ghosh}
\email{sanjay@bosemain.boseinst.ac.in}
\author{\large Tamal K. Mukherjee}
\email{tamal@bosemain.boseinst.ac.in}
\affiliation{Department of Physics, Bose Institute, 93/1, A.P.C Road,
Kolkata - 700 009, INDIA}
\author{\large Munshi G. Mustafa}
\email{munshigolam.mustafa@saha.ac.in}
\author{\large Rajarshi Ray}
\email{rajarshi.ray@saha.ac.in}
\affiliation{Theory Division, Saha Institute of Nuclear Physics,
1/AF, Bidhannagar, Kolkata - 700 064, INDIA}.

\bigskip
\bigskip

\begin{abstract}
\leftskip1.0cm
\rightskip1.0cm
An improved version of the PNJL model is used to calculate various 
thermodynamical quantities, {\it viz.}, quark number susceptibility,
isospin susceptibility, specific heat, speed of sound and conformal 
measure. Comparison with Lattice data is found to be encouraging.
\end{abstract}

\maketitle

\section{Introduction}
Under the
extreme conditions of temperature and/or density a phase transition
is expected in the QCD phase diagram. In fact,
there are two phase transitions at these extreme conditions, namely the
confinement to deconfinement phase
transition and the chiral phase transition. These two phase
transitions are defined in two extreme quark mass
limits. For infinite quark mass limit confinement to deconfinement
phase transition is well defined, and
Polyakov loop acts as the order parameter of the transition corresponding
to the $Z(3)$ (for three colours) global
symmetry breaking. In the limit of zero quark mass we have chiral
symmetry restoring transition. This phase
transition is characterized by the chiral condensate, which acts as
the order parameter. But,
in the real world where quark masses have a finite value, the nature of
these phase transitions is not
known. Whether these two transitions will occur simultaneously or
one will precedes the other remain an
open question. Order of the phase is another aspect of the phase
transition which is also being pursued actively.

Here we are concerned with the thermodynamic aspect of the phase
transition in terms of some relevant
thermodynamic variables such as various quark number and isospin
susceptibilities, specific heat, speed
of sound etc., to see how the behaviour of these variables change
as we go from one phase to the another.
Being in the nonperturbative domain the Lattice QCD and the effective
models are the only theoretical framework which can be employed to
study the phase transition. The effective model we employ here is
known as the PNJL (Polyakov loop + Nambu-Jona-Lasinio) model.
Within the framework of PNJL model one can study both the confinement
to deconfinement transition and the
chiral phase transition together. For details of this model and
of its various improvements  see
Refs.~\cite{pnjl1,pnjl2,pnjl3,ipnjl1,ipnjl2,ipnjl3,ipnjl4,ipnjl5,ipnjl6}

The plan of the paper is as follows. In the next section we
briefly outline the theoretical framework. In
section 3, our results are discussed and finally we conclude
in section 4.


\section{Theoretical Framework}
The coupling between the chiral and the deconfinement order parameter
within the PNJL model allows us to study the thermodynamics of both the
transitions inside a single theoretical framework. The thermodynamics
of the gauge sector is controlled by the Polyakov loop whereas the NJL
model takes care of the thermodynamics of the quark sector.
To obtain various thermodynamic quantities we start with the
thermodynamic potential per unit volume computed within the PNJL
model~\cite{pnjl3,rstm,rsm} :

\begin{eqnarray}
\Omega&=&{\cal U}\left(\Phi,\bar{\Phi},T\right)+
2 G_1(\sigma_u^2 + \sigma_d^2) + 4 G_2 \sigma_u \sigma_d
- \sum_{f=u,d} 6\int\frac{\mathrm{d}^3p}{\left(2\pi\right)^3}{E_f}
\theta\left(\Lambda^2-\vec{p}^{~2}\right) ~~~
\nonumber \\
&-& \sum_{f=u,d}
2\,T\int\frac{\mathrm{d}^3p}{\left(2\pi\right)^3}
 \left\{ \ln\left[1+3\left(\Phi+\bar{\Phi}\mathrm{e}^
{-\left(E_f-\mu_f\right)/T}\right)\mathrm{e}^{-\left(E_f-\mu_f\right)/T}
 + \mathrm{e}^{-3\left(E_f-\mu_f\right)/T}\right]\right . \nonumber\\
&+&
\left .
\ln\left[1+3\left(\bar{\Phi}+\Phi\mathrm{e}^{-\left(E_f+\mu_f\right)/T}
\right)\mathrm{e}^{-\left(E_f+\mu_f\right)/T}+
\mathrm{e}^{-3\left(E_f+\mu_f\right)/T}\right] \right\}
 ~~~,
\end{eqnarray}
where $\sigma_u $ and $\sigma_d$ are, respectively, two light flavours
condensates and the respective
chemical potentials are $\mu_u$ and $\mu_d$. Note that
$\mu_0 = (\mu_u+\mu_d)/2$ and $\mu_I = (\mu_u-\mu_d)/2$. The quasi-particle
energies are $E_{u,d}=\sqrt{\vec{p}^{~2}+m_{u,d}^2}$, where
$m_{u,d}=m_0-4 G_1 \sigma_{u,d} -4 G_2 \sigma_{d,u}$ are the
constituent quark masses and $m_0$ is the current quark mass
(we assume flavour degeneracy).  $G_1$ and $G_2$ are the effective
coupling strengths of a local, chiral symmetric four-point
interaction. We take $G_1 = G_2 = G/4$, where $G$ is the coupling
used in Ref. \cite{pnjl3}.
$\Lambda$ is the 3-momentum cutoff in the NJL model.
The form of the effective gauge potential is in the form given by,
\begin{eqnarray}
\frac{{\cal U}\left(\Phi,\bar{\Phi},T\right)}{T^4} =
\frac{{\cal U^{\prime}}\left(\Phi,\bar{\Phi},T\right)}{T^4}
-\kappa \ln[J(\Phi,\bar{\Phi})] \ ,
\label{effU}
\end{eqnarray}
where ${\cal U^{\prime}}$ is taken from \cite{pnjl3,ipnjl6}.
It arises because of the transformation from
matrix valued field $L$ to complex valued field $\Phi$. The role of this
second term (known as Vandermonde determinant (VdM) term) is to regulate
the behaviour of $\Phi$ so that its value remains within the group
theoretic limit. This term is dropped while calculating the pressure.
For details we refer to \cite{ipnjl6}.
\par
Once the thermodynamic potential $\Omega$ is fixed, we first obtain the
mean fields by minimizing $\Omega$ w.r.t the field variables $\Phi$,
$\bar{\Phi}$, $\sigma_u$ and $\sigma_d$. We then calculate all the relevant
thermodynamic quantities by appropriate differentiation of $\Omega$.
Here we are concerned with the various susceptibilities namely the quark number
susceptibility (QNS) and isospin number susceptibility (INS). It is also
to be noted that by QGP susceptibility we intend to mean these susceptibilities.
Susceptibilities are important as they are related to fluctuations which can
be measured experimentally.

\par
We computed all these quantities from the Taylor expansion of the pressure,
as it is related to the thermodynamical potential via the relation,
\begin{eqnarray}
P(T,\mu_0) = -\Omega(T,\mu_0) ~~~.
\label{prsu}
\end{eqnarray}
For susceptibilities we expand the scaled pressure ($P/T^4$) in a
Taylor series for the quark number and isospin number chemical potentials,
$\mu_0$ and $\mu_I$ respectively. The coefficients we are interested in
are given by,

\begin{eqnarray}
c_n(T) &=& \frac{1}{n!} \left. \frac{\partial^n \left ({P(T,\mu_0) / T^4}
\right ) }
{\partial \left(\frac{\mu_0}{T}\right)^n}\right|_{\mu_0=0} = c^{n0}_n~~~,
\end{eqnarray}

\begin{eqnarray}
c^I_n(T) &=& \left. {\frac{1}{n!} \frac{\partial^n \left ({P(T,\mu_0,\mu_I) /
T^4} \right ) }{
\partial \left(\frac{\mu_0 }{ T }\right)^{n-2}
\partial \left(\frac{\mu_I }{T }\right)^2
}}\right|_{\mu_0=0,\mu_I=0} = c^{(n-2) 2}_n~~~; ~ n > 1.
\end{eqnarray}

The first equation represents the QNS and its higher order derivatives 
whereas the second one corresponds to the INS and its higher order derivatives.
Specific heat and the speed of sound are calculated from the expansion
of the pressure with respect to the temperature.
Explicit expressions of these quantities are given by:

\begin{eqnarray}
C_V = \left . {\partial \epsilon \over \partial T} \right |_V
    = - \left . T {\partial^2 \Omega \over \partial T^2}
\right |_V ~~~,
\label{sph}
\end{eqnarray}

\begin{eqnarray}
v_s^2 = \left . {\partial P \over \partial \epsilon} \right |_S
      = \left . {\partial P \over \partial T} \right |_V \left /
        \left . {\partial \epsilon \over \partial T} \right |_V \right .
      = \left . {\partial \Omega \over \partial T} \right |_V \left /
        \left . T {\partial^2 \Omega \over \partial T^2} \right |_V
        \right .  ~~~.
\label{sps}
\end{eqnarray}
For details of computing these coefficients please refer to \cite{rstm,ipnjl6}.

\section{Result}
\vspace*{-0.3in}
\begin{figure}[!tbh]
\subfigure[]{
\label{fg.c2}
   {\includegraphics [scale=0.50] {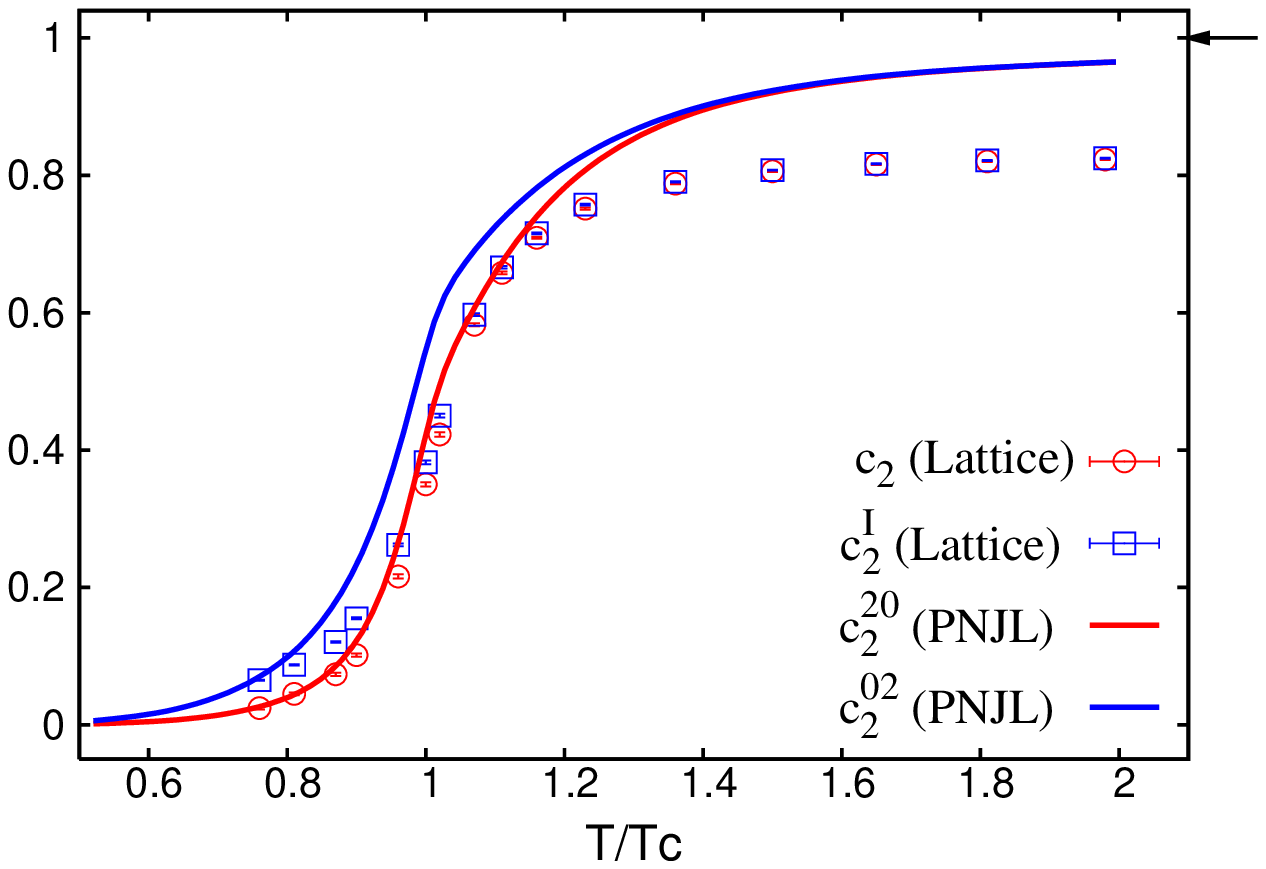}}
}
\hskip 0.1 in
\subfigure[]{
\label{fg.c4}
   {\includegraphics [scale=0.50] {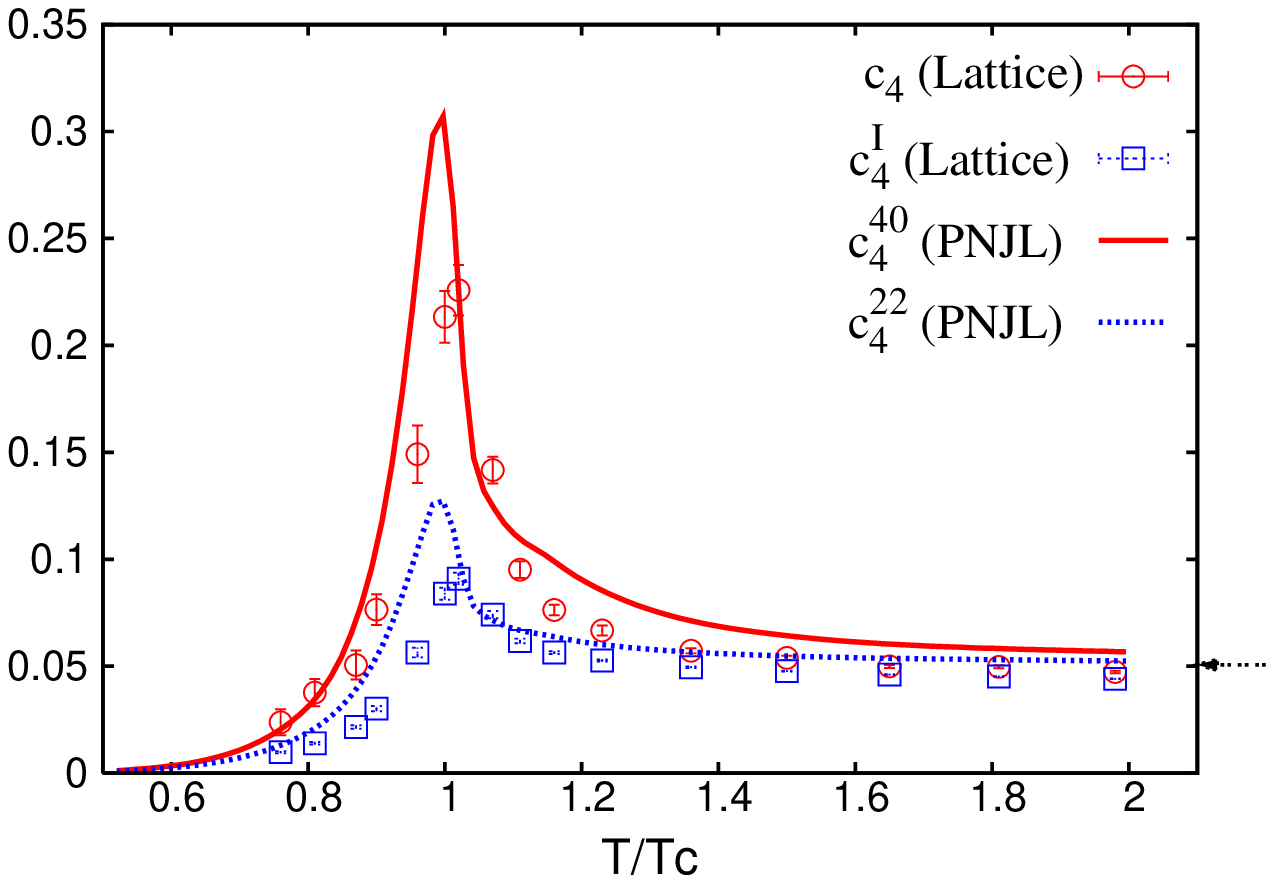}}
}
\vspace*{-0.2in}
   \caption{The Taylor expansion coefficients of the pressure in
            quark number and isospin chemical potentials as functions
            of $T/T_c$. Symbols are LQCD data \cite{lqcd}. Arrows on
            the right indicate the corresponding ideal gas values.
      }
\label{fg.scnord}\end{figure}
We present the QNS, INS and their higher order derivatives with
respect to $\mu_0$ in Fig.\ \ref{fg.scnord}. We have plotted the
LQCD data from Ref.\ \cite{lqcd} for quantitative comparison.
We find, QNS $c_2$ (Fig.\ \ref{fg.c2}) matches well with the
Lattice data upto $1.2 T_c$ but beyond which it goes to SB limit
whereas the Lattice data saturates at $80\%$ of the ideal
gas value. Similar trend is observed for INS $c_2$. It remains
close to LQCD data upto $1.2T_c$ and then it saturates at
SB limit. However both QNS $c_2$ and INS $c_2$
agree well with each other which is also observed in
Lattice calculation. It is to be noted here the QNS $c_2$
matches well with the LQCD data as shown in \cite{rstm},
if we do not include the VDM term.

Variation of next higher order coefficients $c_4$ of both
QNS and INS with temperature are shown in Fig.\ \ref{fg.c4}.
We find a good agreement for both the coefficients with the
Lattice data for whole range of temperatures. If we do not 
include the VDM term (see \cite{rstm}) then the QNS $c_4$
matches the Lattice data only upto $T \sim 1.1 T_c$.
The effect of the VDM term is to bring the $c_2$ and
$c_4$ coefficients to their respective SB limit.

\begin{figure}[!tbh]
\vspace*{-0.3in}
\subfigure[]{
   {\includegraphics [scale=0.50] {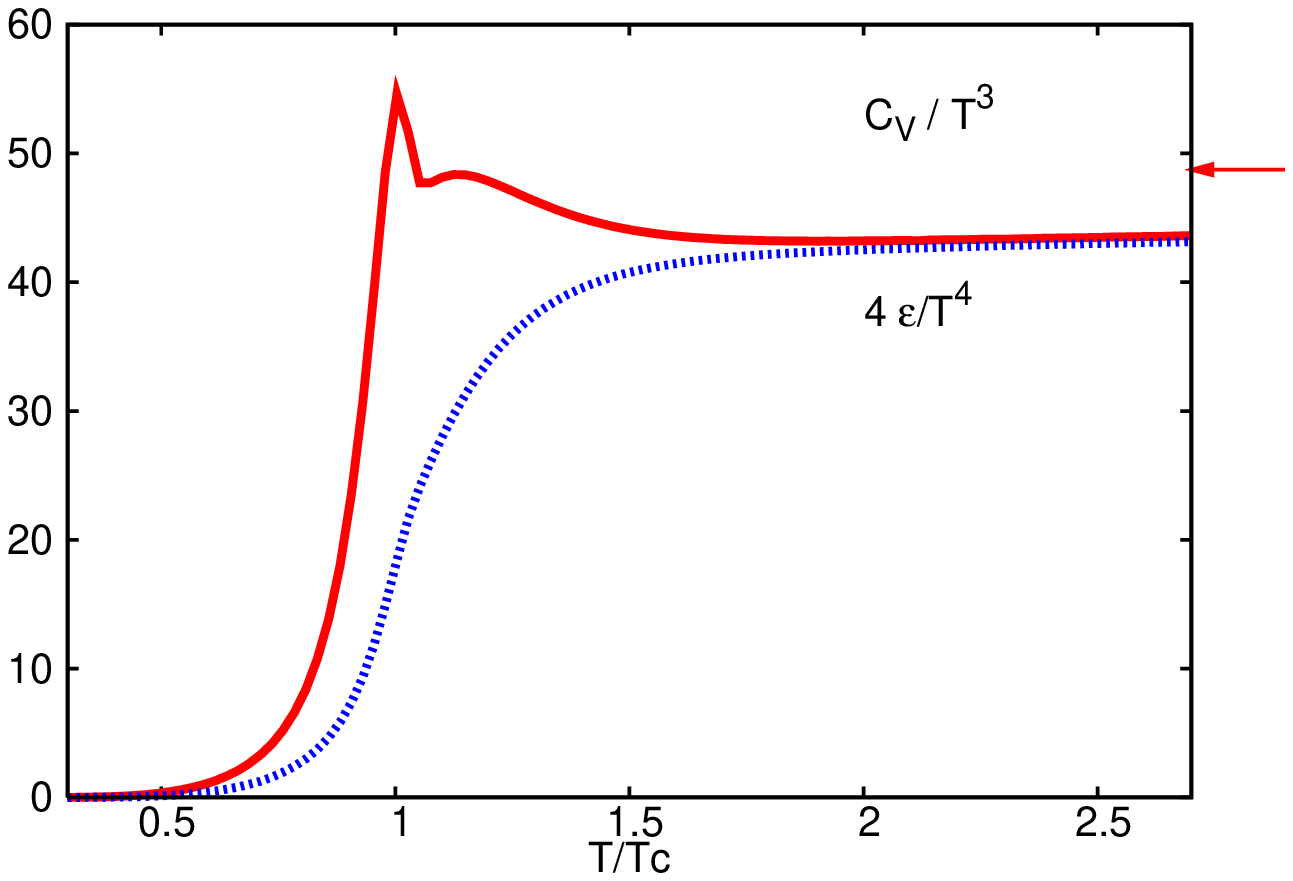}}
\label{fg.cv}
}
\hskip 0.15 in
\subfigure[]{
   {\includegraphics[scale=0.50]{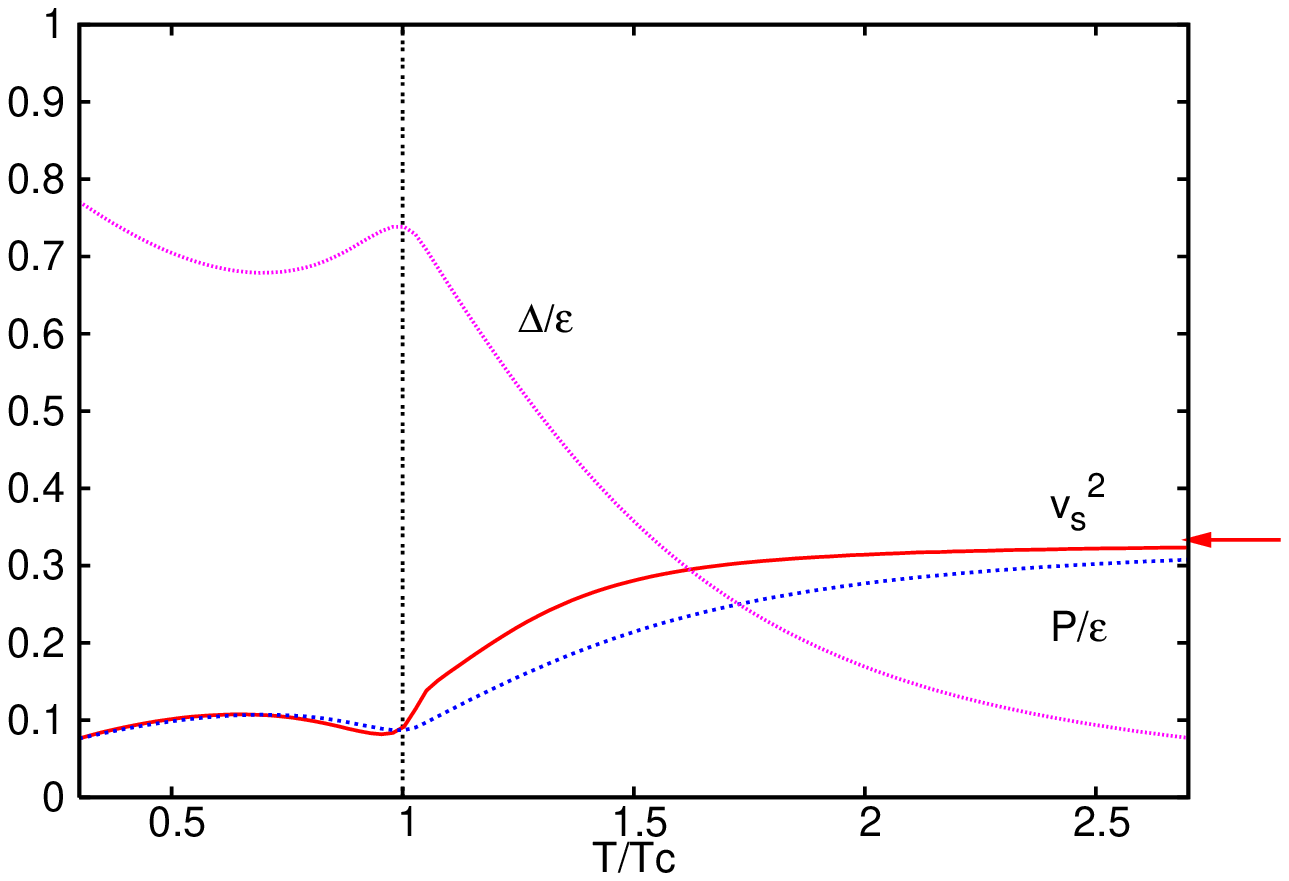}}
\label{fg.cs}
}
\vspace*{-0.2in}
   \caption{ Temperature dependence of  (a): energy density $\epsilon$
            and specific heat $C_V$. (b):
            squared speed of sound $v_s^2$ and conformal measure
            $\Delta/\epsilon$. The arrows in the right show the
            corresponding SB limit.
	   }
\label{fg.cvcs}
\end{figure}

In Fig.~\ref{fg.cvcs} we have presented our result for
specific heat, speed of sound and conformal measure.
Compared to our previous observation in Ref.~\cite{rstm}
the specific heat $C_V$ differs only marginally at
higher temperature.  We find a small peak near the
transition temperature which in the case of critical end point (CEP)
will diverge. At higher temperature it saturates below the ideal
gas limit and approaches the conformal limit of $4 \epsilon/T^4$
from above. The effect of the VDM term cancels out for both
the speed of sound and the conformal measure as they are the
ratios of the energy density and pressure.

\section{Summary}
Various thermodynamical variables of importance are calculated
using an improved version of the PNJL model. All the variables
estimated are found to be in reasonable agreement with the
LQCD data, though the QNS $c_2$ and the INS $c^I_2$ on the Lattice
are smaller by about 20 $\%$.  We found all thermodynamical
variables approach to their respective ideal gas limit at
higher temperature.

\end{document}